\documentclass{llncs}
\usepackage{cite}
\usepackage{graphicx}
\usepackage{amsfonts}
\usepackage[cmex10]{amsmath}
\usepackage{euscript}
\usepackage{pstricks}
\usepackage{color}
\usepackage{flushend}
\usepackage{enumerate}
\usepackage[shortlabels]{enumitem}
\usepackage{subfig}
\usepackage{todonotes}
\usepackage{changes}
\usepackage{algorithm}
\usepackage{algpseudocode}

\begin{document}
\title{Constructing Coverability Graphs for Time Basic Petri Nets}
\author{Matteo Camilli}
\institute{Dept. of Computer Science \\ Universit\`a degli Studi di Milano, Italy \\ \email{camilli@di.unimi.it}}
\maketitle

\normalem

\begin{abstract}
Time-Basic Petri nets, is a powerful formalism for modeling real-time systems where time constraints are expressed through time functions of marking's time description associated with transition, representing possible firing times.
We introduce a technique for coverability analysis based on the building of a finite graph.
This technique further exploits the time anonymous concept \cite{Bellettini11,2014arXiv1409.2778C}, in order to deal with topologically unbounded nets, exploits the concept of a coverage of $TA$ tokens, i.e., a sort of $\omega$
anonymous timestamp.
Such a coverability analysis technique is able to construct coverability trees/graphs for unbounded Time-Basic Petri net models. The termination of the algorithm is guaranteed as long as, within the input model, tokens growing without limit, can be anonymized. This means that we are able to manage models that do not exhibit Zeno behavior and do not express actions depending on ÒinfiniteÓ past events. This is actually a reasonable limitation because, generally, real-world examples
do not exhibit such a behavior.

\end{abstract}

\section{Introduction}
\label{sec:tbcg}
When analyzing a Petri net, a very common question is whether or not the net is bounded. If it is bounded,
the net is theoretically analyzable, and its state space is finite. However the net may be unbounded and classic 
state space methods generates an infinite number of reachable states from these kind of models. 
Time Basic (TB) Petri nets \cite{UnifiedWay91}, as classic Place/Transition nets, may be topologically unbounded.
The unboundedness happens whenever there exists a place in the net, where it is possible to accumulate an infinite number of tokens during its execution.
Coverability graph algorithms are able to deal with such a models and allow us to decide several important problems: the \emph{boundedness} problem (BP), 
the \emph{place-boundedness} problem (PBP), the \emph{semi-liveness} problem (SLP) and the \emph{regularity} problem (RP) \cite{Karp69,Valk81}.
Anyway, for TB nets, this task is complicated by the time domain. In fact,
tokens come along with temporal information and, in general, it is not possible to cluster them into an $\omega$ symbol without loosing important information about the system's behavior.
However, a technique able to construct a finite symbolic reachability graph ($TRG$) relying on a sort of time coverage, was recently introduced \cite{Bellettini11,2014arXiv1409.2778C}
This technique overcomes the limitations of the existing available analyzers for TB nets,
based in turn on a time-bounded inspection of a (possibly infinite) reachability-tree.
The time anonymous concept \cite{Bellettini11,2014arXiv1409.2778C}, introduced by such a technique,
allow us to overcome the issue of clustering tokens. In fact, time anonymous timestamps do not carry, for definition, any temporal information. Therefore, an infinite number of $TA$ tokens can be clustered together into a $TA^\omega$ symbol
without loss of information. The technique, introduced in the current report, gives us a means to deal with 
topologically unbounded TB net models, where the unboundedness refers to places having an infinite number of $TA$
tokens.
Such a limitation is actually reasonable, in practice.
In fact, this restricts the analyzable models to systems which do not exhibit \emph{Zeno} behavior and
do not express actions depending on ``infinite'' past events.

\begin{figure}[htbf]
\centering
\includegraphics[width=.5\textwidth]{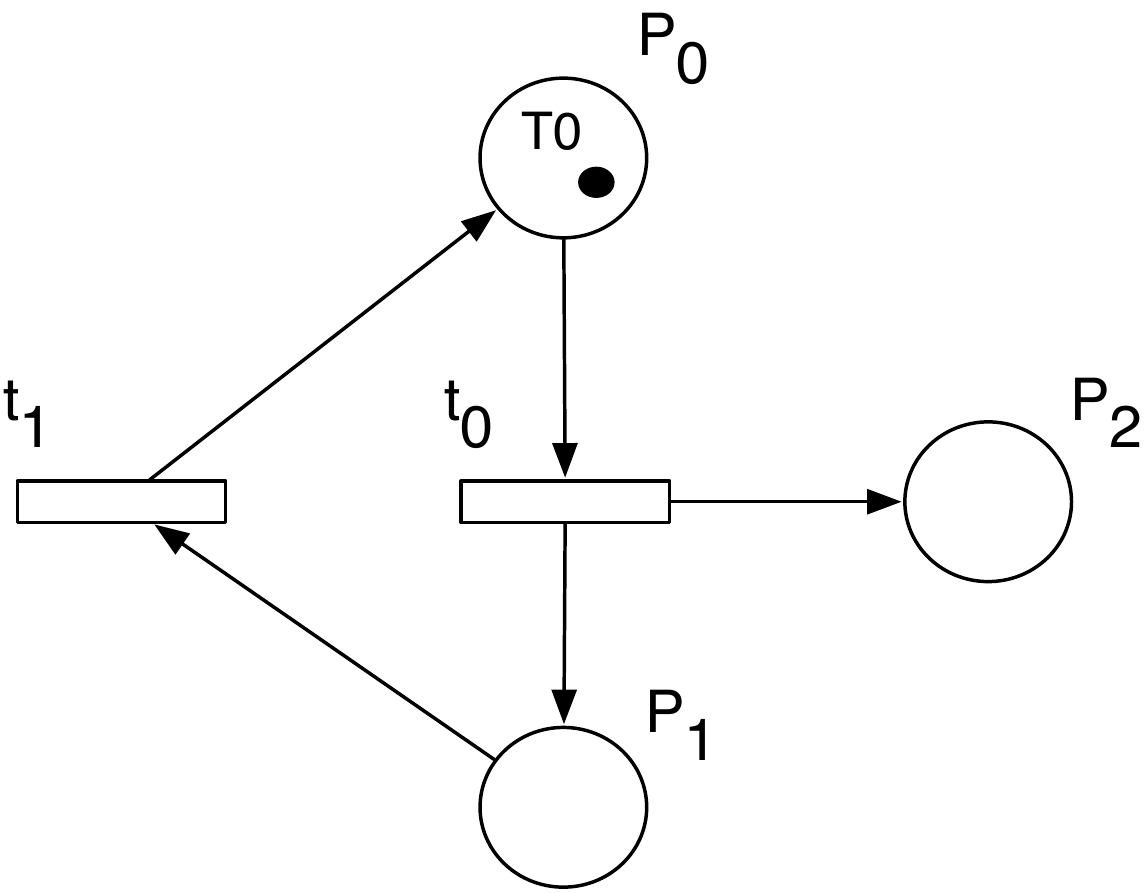}
{\small
\vskip 0.5cm
{\scriptsize
\begin{tabular}{@{} llll @{}}
\bf{Initial marking} &  $P_0\{T_0\}$ \\
\bf{Initial constraint} & $T_0 \ge 0$ \\
 \end{tabular}
 \\
 \begin{tabular}{@{} llll @{}}
\bf{$t_0$} & [$enab + 1.0 , enab + 2.0$] \\
\bf{$t_1$} & [$enab + 1.0, enab + 2.0$] \\
\end{tabular}
}}
\caption{Simple example showing an unbounded TB net model.}
\label{fig:TW-es1}
\end{figure}

\begin{figure}[htbf]
\centering
\includegraphics[width=\textwidth]{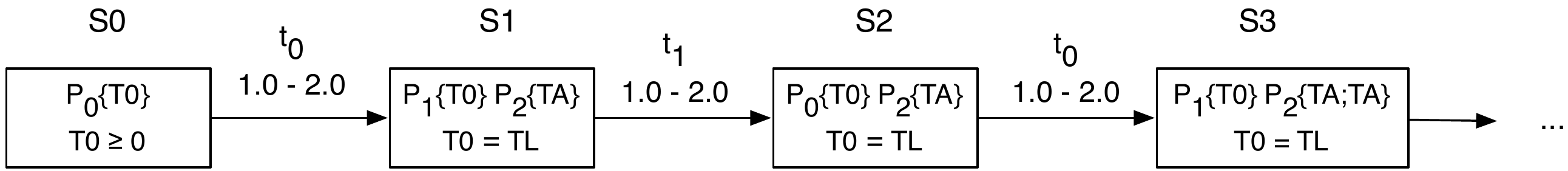}
\caption{Portion of the infinite reachability tree associated to the TB net model presented in Figure \ref{fig:TW-es1}.}
\label{fig:TW-es1-graph}
\end{figure}

As a simple example, consider the model described in Figure \ref{fig:TW-es1}.
The behavior of the system is very simple: from the initial state, the transition $t_0$ must fire in the time interval $[T0, T0+2.0]$. Its firing consumes $T0$ and produces two new tokens in places $P_1$ and $P_2$, respectively.
In this new state, $t_1$ is the only enabled transition, and its firing brings the system in the initial topological marking.
It is worth noting that every time $T_0$ fires, a new token is placed into $P_2$ which cannot be consumed by any firing transition.
Therefore, the abstraction technique introduced in \cite{Bellettini11,2014arXiv1409.2778C} applied to this example, generates an infinite number of reachable symbolic states because the number of tokens in place $P_2$ grows without limit.
Figure \ref{fig:TW-es1-graph} shows a portion of the infinite reachability tree.

As we can see, the number of $TA$ tokens in place $P_2$ grows indefinitely, thus the execution of the software tool \textsc{Graphgen} \cite{Bellettini11,2014arXiv1409.2778C}, on such a input, does not terminate.
The current report, introduces an extension of the previous analysis technique able to build the coverability graph
of unbounded TB nets, 
exploiting the concept of $TA$ coverage tokens.
Our proposal takes inspiration from the Monotone-Pruning (MP) algorithm introduced in \cite{Reynier11},
for P/T nets, and extends it to deal with TB net models, thus supplying a means,
also for real-time systems, 
to solve the above mentioned problems.

\subsection{Preliminaries}
\label{sec:pre}
A quasi order $\geq$ on a set $S$ is a reflexive and transitive relation on $S$.
Given a quasi order $\geq$ on $S$, a state $s \in S$ and a subset $X$ of $S$, we write 
$s \geq X$ iff there exists an element $s' \in X$ such that $s \geq s'$.

Given a finite set of places $P$, the marking $M$ (\cite{Bellettini11,2014arXiv1409.2778C}) on $P$ 
is a function $P \rightarrow Bag(T S \cup \{TA\})$ which supplies foreach place, timestamps associated with
tokens.
The symbolic $\omega$-marking $M^\omega$ on $P$ is a function $P \rightarrow Bag(T S \cup \{TA, TA^\omega \})$. 
The $TA^\omega$ symbol represents, in this case, any number of $TA$ symbols ($\infty$ included).
Given the set $U(P) = \mathbb{N}^{|P|}$, an $u$-marking $\bar{u}$, 
is an element of $U(P)$ which associates foreach place,
the number of non-$TA$ tokens.
Given the set $V(P) = (\mathbb{N} \cup \{\omega\})^{|P|}$, an $v$-marking $\bar{v}$, 
is an element of $V(P)$ which associates foreach place,
the number of $TA$ tokens.
Given a symbolic state $S$, we denote with $\bar{u}(S)$, and $\bar{v}(S)$ the $u$-marking and the $v$-marking
associated with $S$, respectively.  

Given an element $\bar{u} \in U(P)$, $\bar{v} \in V(P)$, and a place $p$, we denote with 
$\bar{u}_p$ the number of non-$TA$ tokens in place $p$, and with
$\bar{v}_p$ the number of $TA$ tokens in place $p$.
Since the $\omega$ symbol represents an infinite number of $TA$ tokens,
the component $\bar{v}_p = \omega$ if and only if $TA^\omega \in M^\omega(p)$.


For instance, if $P=\{p_1, p_2, p_3, p_4\}$ and the symbolic $\omega$-marking is $\{ p_1\{T0, TA\}$, $p_3\{T0, T1, TA^\omega \} \}$, the corresponding $u$-marking, and $v$-marking
are $\{ 1,0,2,0\}$, and $\{ 1,0,\omega,0 \}$, respectively.

The set $V(P)$ is equipped with a partial order $\geq$ naturally extended by letting $n < \omega, \forall n \in \mathbb{N}$ and $\omega \geq \omega$.

In the current report, when referring to symbolic states, we consider an extended version of the definition 
proposed in \cite{Bellettini11,2014arXiv1409.2778C}, where the marking is represented by the function $M^\omega$ rather than $M$.

\begin{definition}[TA erasure]
\label{def:ta-erasure}
Given a symbolic state $S = \langle M^\omega, C \rangle$, $S_{[\neg TA]}$ is a symbolic state composed of
$\langle {M^\omega}', C \rangle$,
where ${M^\omega}'$ is a symbolic $\omega$-marking obtained from the erasure of all $TA$ symbols from $M^\omega$.
\end{definition}

\begin{definition}[state coverage]
\label{def:symb-state-covering}
Given two symbolic states $S = \langle M^\omega, C \rangle$ and $S' = \langle {M^\omega}', C' \rangle$,
the $u$- and $v$- markings of $S$ $\bar{u}$, $\bar{v}$, and 
the $u$- and $v$- markings of $S'$ $\bar{u}'$, $\bar{v}'$,
$S$ covers $S'$ ($S \geq S' $) iff $\bar{u} = \bar{u}' \wedge \bar{v} \geq \bar{v}' \wedge C \equiv C'$.
\end{definition}

That means that $S$ differs from $S'$ only in the number of $TA$ tokens in places. In particular,
the number of $TA$ tokens foreach place in $S$ is greater or equal to those ones foreach place in $S'$.
Formally, $\forall p \in P, \bar{v}_p \geq \bar{v}'_p$.
Whenever $S \geq S'$ and $\bar{v} \neq \bar{v}'$ we say that $S$ properly covers $S'$, and we denote it with
$S > S'$.

\begin{definition}[Coverability Tree]
\label{def:cover-tree}
Given a TB net $\mathcal{R} = \langle P, T, F \rangle$, 
a coverability tree is a tuple $\mathcal{T}=\langle N, n_0, E \rangle$, where $N$ is a set of symbolic states, $n_0 \in N$ is the toot state, $E \subseteq N \times T \times N$ is the set of edges labeled with firing transitions. Where:
\begin{enumerate}
\item foreach reachable symbolic state $S$ in $TRG(\mathcal{R})$, there exists $S' \in N$ s.t. either $S \subseteq S'$ or $S \geq S'$.
\item foreach symbolic state $S= \langle M^\omega, C \rangle \in N$, having $u$-marking $\bar{u}$
and $v$-marking $\bar{v}$, there exists either a reachable state $s$ of $\mathcal{R}$ s.t. $s \in S$, or a an infinite sequence of reachable symbolic states in $TRG(\mathcal{R})$, $(S_n)_{n\in\mathbb{N}}$ s.t.
$\forall n, C_n \equiv C$ and 
$\forall n$, $\bar{u}(S_n) = \bar{u}$ and 
the sequence $(\bar{v}(S_n))_{n\in\mathbb{N}}$ is strictly increasing converging to $\bar{v}$.
\end{enumerate}
\end{definition}
Given a symbolic state $S \in N$, we denote by \emph{Ancestor}$_{\mathcal{T}}(S)$ the set of ancestors of $S$ in
$\mathcal{T}$ ($S$ included). If $S$ is not the root of $\mathcal{T}$, we denote by
\emph{parent}$_{\mathcal{T}}(S)$ its first ancestor in $\mathcal{T}$.
Finally, given two symbolic states $S$ and $S'$ such that $S \in$
\emph{Ancestor}$_{\mathcal{T}}(S')$, we denote by \emph{path}$_{\mathcal{T}}(S, S') \in E^*$ the sequence of edges leading from $S$ to $S'$ in $\mathcal{T}$. 

\subsection{Coverability Tree Algorithm}
\label{sec:cg-alg}

This section presents the algorithm able to construct coverability trees of TB nets.
We call it
$TBCT$ (Algorithm \ref{alg:tbct}) and it is inspired by the Monotonic pruning (MP) algorithm introduced in \cite{Reynier11}, able to build minimal 
coverability sets for classic P/T nets.
Our proposal involves the acceleration function \emph{Acc}, first introduced in the Karp and Miller algorithm \cite{Karp69}. However, it is defined and also applied in a slightly different manner, in order to deal with a different modeling formalism.
In the current context, the \emph{Acc} function, actually modifies the symbolic $\omega$-marking
$M^\omega$ of a symbolic state by inserting proper $TA^\omega$ symbols, accordingly to the following:
$$Acc : 2^N \times N \rightarrow N, Acc(W, S)(p) = S' \text{ s.t.}$$
\begin{equation}
\label{eq:acc}
\forall p \in P, \bar{v}(S')_p = \left \{ 
  \begin{array}{l l}
    \omega & \quad \text{if $\exists S'' \in W : S'' < S \wedge \bar{v}(S'')_p < \bar{v}(S)_p \wedge S'' \rhd S$ }\\
    m_{S}(p)_2 & \quad \text{otherwise}
  \end{array} \right.
\end{equation}

Where $S'' \rhd S$ iff there exists $\sigma = $\emph{path}$_{\mathcal{T}}(S'', S)$, such that $\sigma$ is feasible from $S$.
Such a condition holds whenever, either:
\begin{enumerate}
\item $C_{S''} \implies C_{S}$, meaning that, ${S''}_{[\neg TA]} \subseteq S_{[\neg TA]}$. In this case,
all the paths starting from $S''$ are feasible from $S$.
\item $C_{S} \implies C_{S''}$ and the first component of $\sigma$ is of type \texttt{A*} \cite{Camilli13}. In this case ${S}_{[\neg TA]} \subseteq {S''}_{[\neg TA]}$, therefore not all paths starting from $S''$ are feasible from $S$, but since $\sigma$ starts from all ordinary states of $S''$, $\sigma$ is feasible also from $S$. 
\end{enumerate}

For instance, considering the example in Figure \ref{fig:TW-es1-graph}, the evaluation of the \emph{Acc} function on $S2$ and its ancestors: \emph{Acc(\{S0, S1\}, S2)}, causes the insertion of the $TA^\omega$ symbol into $P_2$ because $S2 > S0$, $\bar{v}(S2)_{P_2} > \bar{v}(S0)_{P_2}$ and the path from $S0$ to $S2$ is feasible from $S2$.
This way, we recognize that $TA$ tokens into place $P_2$ can grow without limit.

\begin{algorithm}[h!]
\caption{TBCT Algorithm.}
\label{alg:tbct}
\begin{algorithmic}[1]
\Require{A TB net $\mathcal{R} = \langle P,T,F \rangle$}
\Ensure{A coverability tree $\mathcal{T} = \langle N, n_0, E, L \rangle$, $N = Act \cup Inact$}
\Function{TBCT}{$\mathcal{R}$}
\State $r = BuildRoot(\mathcal{R})$
\State{$N = \{r\}$; $Act = N$; $Wait = N$; $E = \emptyset$; $L = \emptyset;$}
\While{$ Wait \neq \emptyset$}
\State{$s = Pop(Wait)$;}
\If{$s \in Act$} \label{line:act-test}
\For{$t \in EnabledTransitions(s, \mathcal{R})$}
\State $m = Successor(s, t)$; \label{line:succ}
\State $n = Acc(Ancestors_{\mathcal{T}}(m) \cap Act, m)$; \label{line:acc}
\State $N += \{n\}; ~E += \{\langle s, t, n \rangle \};$
\If{$\not\exists a \in Act : a \supseteq n$} \label{line:test-included}
\If{$\exists a \in Act : a \subset n$} \label{line:test-includes}
\State $Act -= \{x : a \in Ancestors_{\mathcal{T}}(x) \}$; \label{line:includes}
\EndIf
\If{$\not\exists a \in Act : a \geq n$} \label{line:covered}
\State $Act -= \{ x : \exists y \in Ancestors_{\mathcal{T}}(x) \text{ s.t } y \leq x$ \\
\indent \indent \indent \indent \indent \indent \indent $\wedge ~ (y \in Act \vee y \in Ancestors_{\mathcal{T}}(n)) \};$ \label{line:deactivation}
\State $Act += \{n\}; ~ Wait += \{n\}$ \label{line:covered-2}
\EndIf
\EndIf
\EndFor
\EndIf
\EndWhile
\EndFunction
\end{algorithmic}
\end{algorithm}

Likewise both the Karp and Miller and the MP Algorithms, the $TBCT$ algorithm builds a coverability tree, but
nodes, in the current context, are symbolic states containing symbolic $\omega$-markings and edges are labeled by transitions of the
analyzed TB net. Therefore it proceeds by exploring the reachability tree of the net \cite{Bellettini11,2014arXiv1409.2778C}, and accelerating along
branches to reach ``limit'' symbolic $\omega$-markings (containing proper $TA^\omega$ symbols).
In addition, during the exploration, it can prune branches that are covered by nodes on other branches.
Therefore, nodes of the tree are partitioned in two subsets: \emph{active} nodes, and \emph{inactive} ones. Intuitively, active nodes will form the coverability set of the TB net, while inactive ones are not part of the final coverability set, because they are dominated by other active nodes.

The Algorithm \ref{alg:tbct} proceeds in the following steps to 
decide how to change the structure $\mathcal{T}$ according to new computed reachable symbolic states:
\begin{enumerate}
\item The symbolic state $s$, popped from $Wait$ should be active (test of Line \ref{line:act-test}).
\item The algorithm iterates through all the enabled transitions and computes one by one all the successor symbolic states: $m = Successor(s, t);$ (Line \ref{line:succ}).
\item The state $m$ is accelerated w.r.t. its active ancestors. A new symbolic state $n$ is created by this operation:
$n = Acc(Ancestors_{\mathcal{T}}(m) \cap Act, m);$ (Line \ref{line:acc}).
\item If the new symbolic state $n$ is not included or equal to one of the existing active nodes, then it is candidate to be active (test of Line \ref{line:test-included}).
\item If the new symbolic state $n$ includes an existing active node $a$, then the sub-tree with root $a$ is deactivated (Lines \ref{line:test-includes}-\ref{line:includes}).
\item The new symbolic state $n$ is declared as active iff it is not covered by any existing active nodes
(test of Line \ref{line:covered} and Line \ref{line:covered-2}).
\item If $n$ is not covered, some symbolic states are deactivated (Line \ref{line:deactivation}).
\end{enumerate}

\begin{figure}
    \centering
    \subfloat[$y \notin Ancestors_{\mathcal{T}}(n)$]{{\includegraphics[height=5cm]{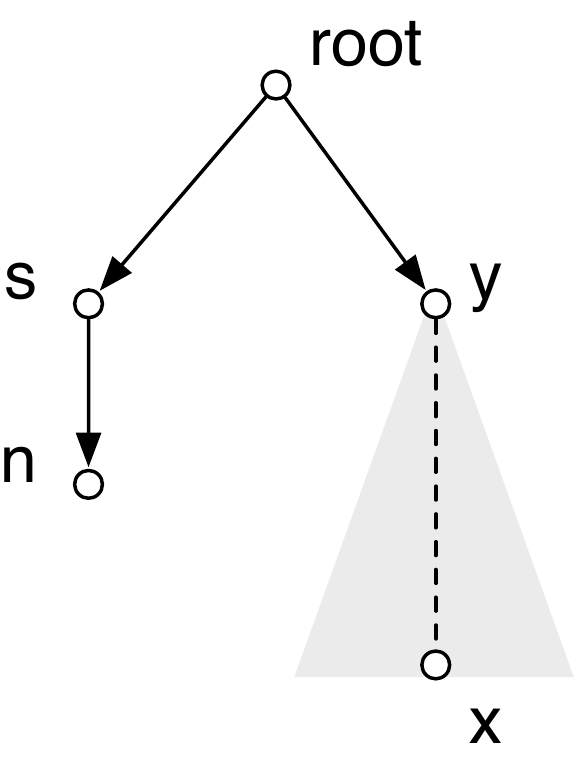} } \label{fig:acc1} } 
    \qquad
    \subfloat[$y \in Ancestors_{\mathcal{T}}(n)$]{{\includegraphics[height=5.8cm]{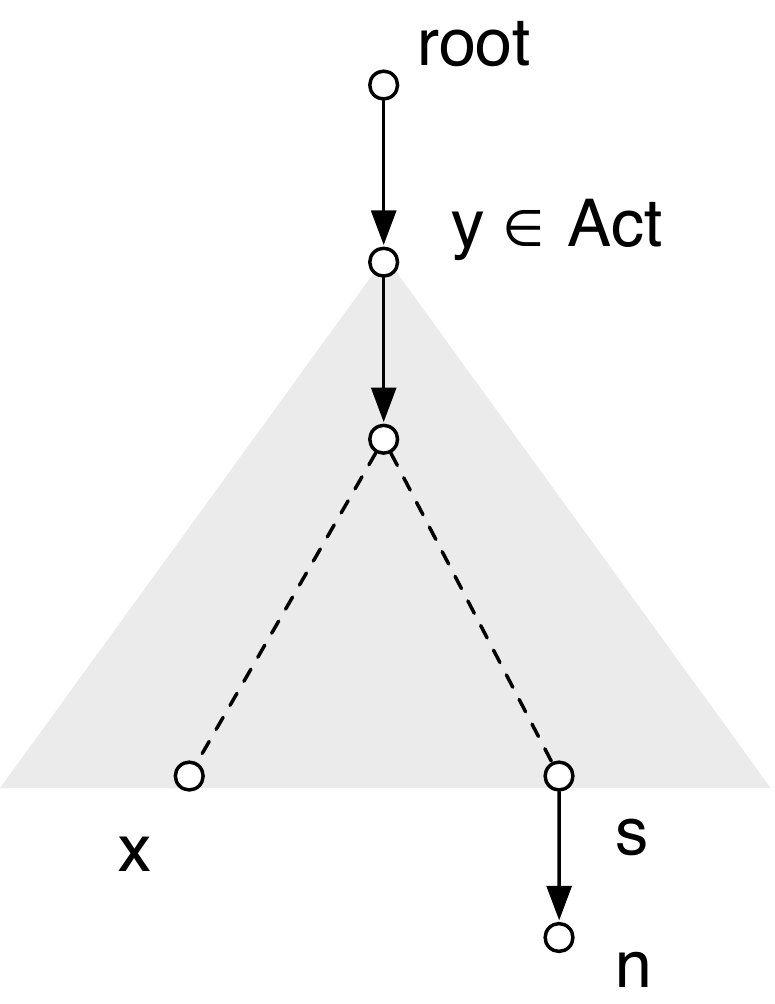} } \label{fig:acc2} }
    \caption{Deactivations of symbolic states in the TBCT Algorithm}
    \label{fig:example}
\end{figure}

The update of the set \emph{Act}, complies with the following rules. Intuitively, nodes (and their descendants)
are deactivated if they are included or covered by other nodes. This would lead to deactivate a node $x$ iff it owns
an ancestor $y$ dominated by $n$, i.e. such that either $y \subset n$ (Lines \ref{line:test-includes}-
\ref{line:includes} ) or $y \leq n$ (\ref{line:covered}-\ref{line:deactivation}). 
Concerning the latter case, whenever a new node $n$ (obtained from \emph{Wait}) covers a node $y$ ($y \leq n$),
then $y$ can be used to deactivate nodes in two ways:
\begin{itemize}
\item if $y \notin Ancestors_{\mathcal{T}}(n)$, then no matter whether $y$ is active or not, all its descendants are deactivated (Figure \ref{fig:acc1}).
\item if $y \in Ancestors_{\mathcal{T}}(n)$, then $y$ must be active ($y \in Act$), and in that case all its descendants are deactivated, except node $n$ itself as it is added to \emph{Act} (Line \ref{line:covered-2}). We require $y \in Act$ to avoid useless
operations. In fact, descendants of $n$ dominates descendants of $y$ (Figure \ref{fig:acc2}).
\end{itemize}

For example, con sidering the example in Figure \ref{fig:TW-es1-graph}, the insertion of $S2$ accelerated
causes the deactivation of both $S0$ and $S1$ because of the
execution of line \ref{line:deactivation}. In particular, such a situation corresponds to Figure \ref{fig:acc2}, because
$S2 \geq S0$ and $S0$ (\emph{active} node) belongs to \emph{Ancestors}$_{\mathcal{T}}(S2)$.

\begin{figure}[htbf]
\centering
\includegraphics[width=\textwidth]{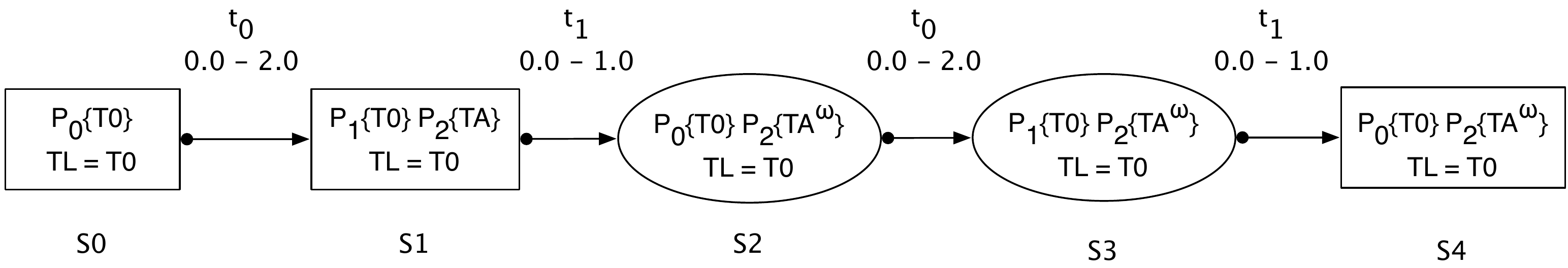}
\caption{Coverability tree constructed from the TB net example presented in Figure \ref{fig:TW-es1}.}
\label{fig:tw-es1-tree}
\end{figure}

Figure \ref{fig:tw-es1-tree} depicts the coverability tree constructed from the TB net example presented in Figure \ref{fig:TW-es1}. Elliptic symbolic states form the final coverability set (\emph{active} nodes), while the squared ones
are symbolic states deactivated during the analysis. As we can see, the $TBCT$ algorithm builds a finite tree
structure from an
unbounded TB net model. In particular, as shown before, the algorithm is able to identify that the number of $TA$
tokens in place $P_2$ can grow without limit.

As we can see in Figure \ref{fig:tw-es1-tree}, edges carry information about their type
(either \texttt{AA}, \texttt{EE}, \texttt{AE} or \texttt{EA} \cite{Camilli13}), and about
the local minimum-maximum firing time.
In the following, given an edge $e$, we refer to these information with 
\emph{type(e)} and \emph{time(e)}, respectively. In particular we refer to the source type with
\emph{type(e)}$_{src}$ and to the target type with \emph{type(e)}$_{trgt}$.

It is also possible to construct a coverability graph $\mathcal{G}$ rather than a tree.
This task, starting from the tree structure $\mathcal{T} = \langle N, n_0, E \rangle$, 
executes the following steps:
\begin{enumerate}
\item All \emph{inactive} nodes are erased from $N$.
\item $\forall a \in Act, \forall \langle a, t, b \rangle \in E$, if $b$ is \emph{inactive}, we search for $a' \in Act$ so that
$a' \supseteq b$ or $a' \geq b$, thus we remove $ \langle a, t, b \rangle$ from $E$ and we insert $\langle a, t, a' \rangle$.
\item All covered edges (Definition \ref{def:edge-covering}) are removed from $E$.
\end{enumerate}

\begin{definition}[edge coverage]
\label{def:edge-covering}
Given a coverability tree $\mathcal{T} = \langle N, n_0, E \rangle$ and two edges $e = \langle a, t, b \rangle $,
$e' = \langle a', t', b' \rangle$ $\in E$,
$e$ covers ($\geq$) $e'$ iff:
\begin{enumerate}[i]
\item $a = a' \wedge t = t' \wedge b = b'$
\item \emph{time}$(e) \supseteq$ \emph{time}$(e')$
\item \emph{type(e)}$_{src} \geq$  \emph{type(e')}$_{src}  \wedge$
\emph{type(e)}$_{trgt} \geq$  \emph{type(e')}$_{trgt}$, being \texttt{A} $>$ \texttt{E}
\end{enumerate}
\end{definition}

\begin{figure}[htbf]
\centering
\includegraphics[width=.4\textwidth]{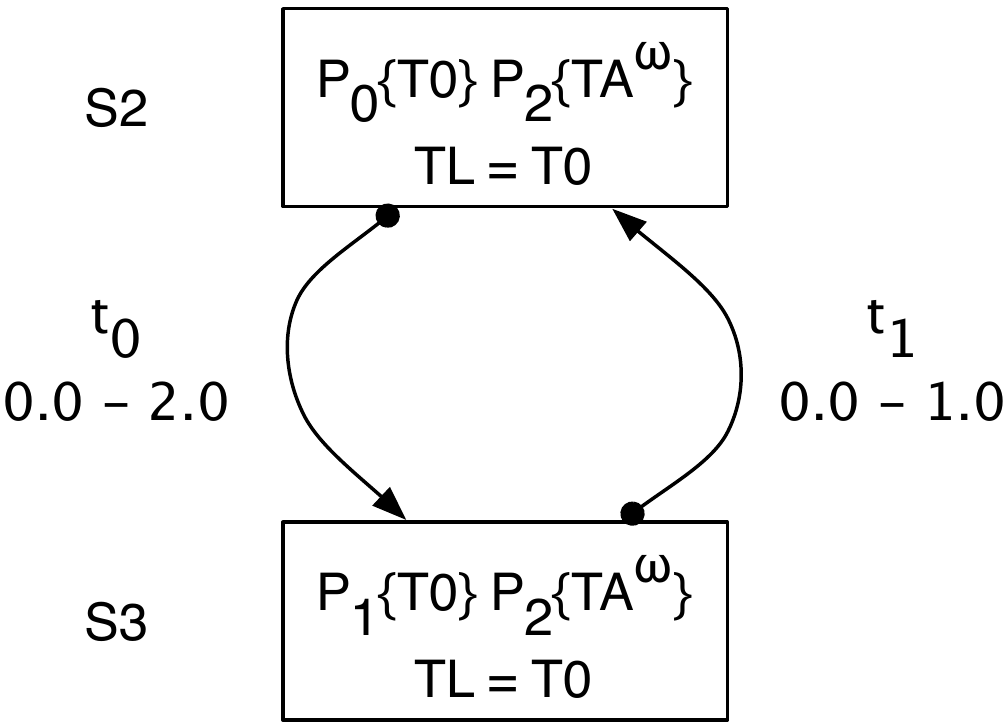}
\caption{Coverability graph constructed from the coverability tree presented in Figure \ref{fig:tw-es1-tree}.}
\label{fig:tw-es1-graph}
\end{figure}

Figure \ref{fig:tw-es1-graph} shows the coverability graph resulting from the coverbility tree presented in figure \ref{fig:tw-es1-tree}. Such a structure contains only active symbolic states and gives us a more intuitive overview on the system's behavior. For instance, by observing Figure \ref{fig:tw-es1-graph}, it's easy to figure out that the system
alternates two symbolic states where $P_0$ and $P_1$ are marked with a single token, while place $P_2$ can
accumulate $TA$ tokens without limit.

The rest of this section reports some simple examples of unbounded TB net models analyzed by the software tool
implementing the $TBCT$ algorithm. All the coverability trees/graphs have been automatically obtained by using \textsc{GraphViz} visualization software \cite{graphviz} on the output generated from the tool-set.
The $TW$ notation used into symbolic $\omega$-markings, stands for $TA^\omega$.

\begin{figure}[htbf]
\centering
\includegraphics[width=.3\textwidth]{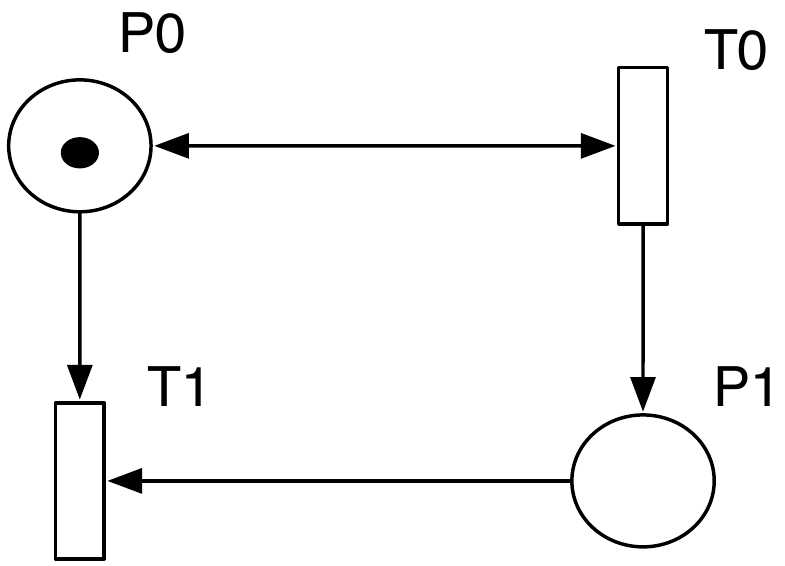}
{\small
\vskip 0.5cm
{\scriptsize
\begin{tabular}{@{} llll @{}}
\bf{Initial marking} &  $P0\{T0\}$ \\
\bf{Initial constraint} & $T0 \ge 0$ \\
 \end{tabular}
 \\
 \begin{tabular}{@{} llll @{}}
\bf{$T0$} & [$enab + 1.0 , enab + 1.0$] \\
\bf{$T1$} & [$enab + 1.0, enab + 1.0$] \\
\end{tabular}
}}
\caption{Unbounded TB net model A.}
\label{fig:tw-esA}
\end{figure}

\paragraph{Example A}
Figure \ref{fig:tw-esA} depicts an unbounded TB net model with two places ($P0$, $P1$)
and two transitions ($T0$, $T1$).
It represents a simple synchronous system, where an operation occurs at each time unit
(e.g. production/consumption). Produced units are stored into a infinite buffer. 
After the first consumption the system stops.

\begin{figure}
    \centering
    \subfloat[Coverability tree of A]{{\includegraphics[width=.4\textwidth]{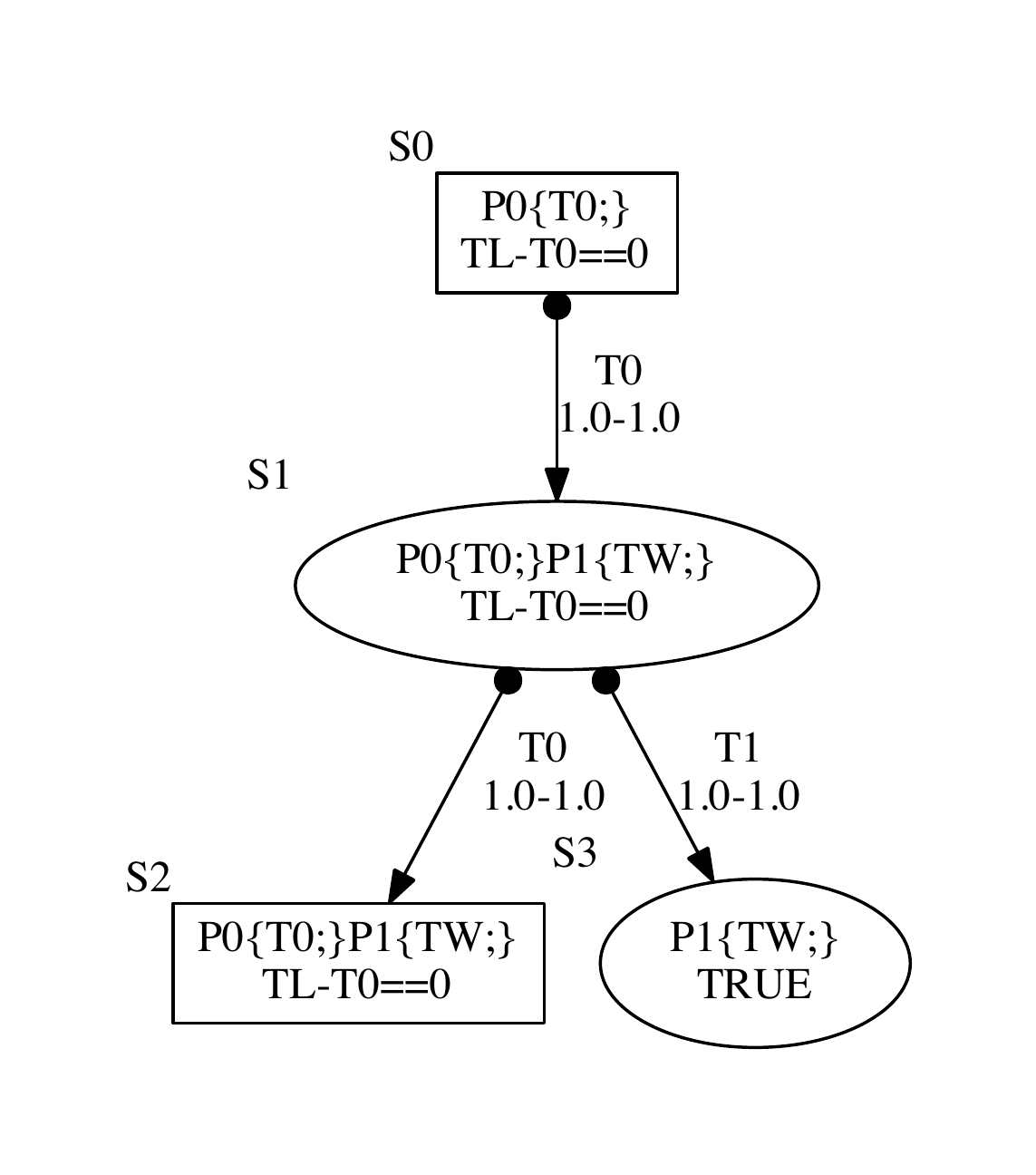} } \label{fig:A-tree} }
     \qquad
    \subfloat[Coverability graph of A]{{\includegraphics[width=.4\textwidth]{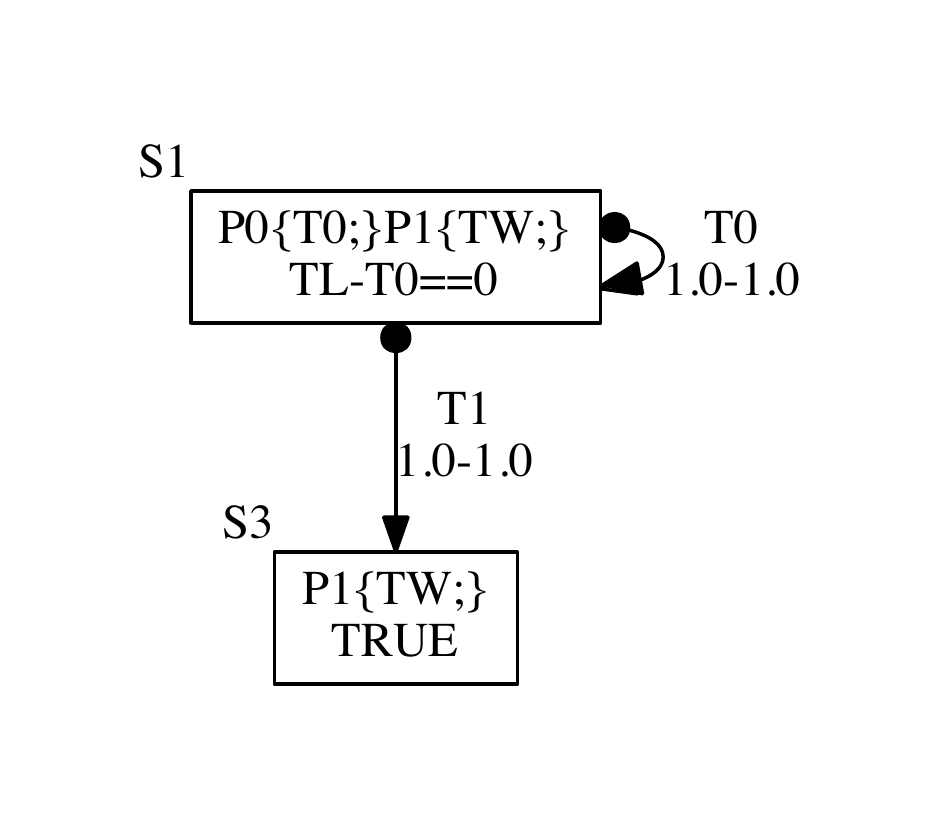} } \label{fig:A-graph} }
    \caption{Coverability tree/graph of example A.}
    \label{fig:tw-esA-tree-graph}
\end{figure}

Figure \ref{fig:A-tree} shows the coverability tree of $A$. As we can see, the introduction of $S1$ causes the deactivation of $S0$ ($S1 \geq S0$). From $S1$ the system can evolve either into $S2$ which is inactive ($S2 = S1$), or $S3$ which is a final state.
Such a behavior is also shown by the coverability graph (Figure \ref{fig:A-graph}): the system loops into $S1$, 
by the firing of $T0$ transition, until the firing of $T1$ which leads into the final state $S3$.

\begin{figure}[htbf]
\centering
\includegraphics[width=.3\textwidth]{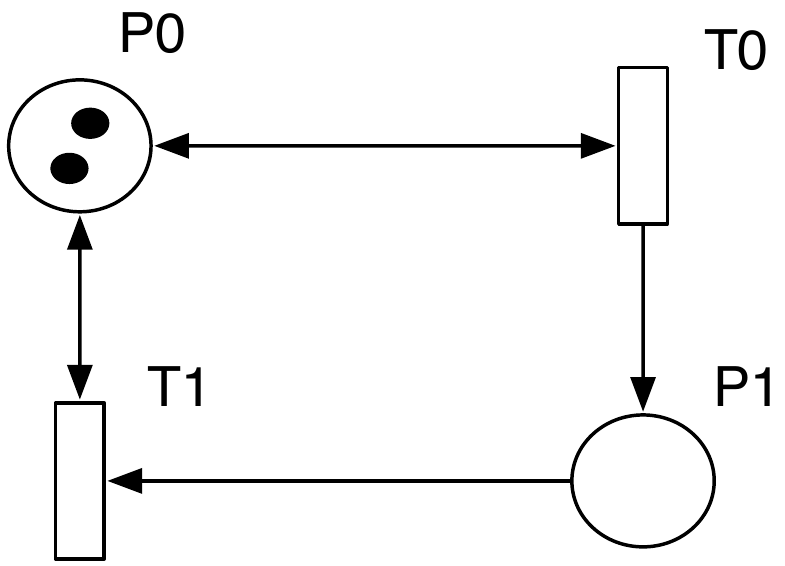}
{\small
\vskip 0.5cm
{\scriptsize
\begin{tabular}{@{} llll @{}}
\bf{Initial marking} &  $P0\{T0,T0\}$ \\
\bf{Initial constraint} & $T0 \ge 0$ \\
 \end{tabular}
 \\
 \begin{tabular}{@{} llll @{}}
\bf{$T0$} & [$enab + 1.0 , enab + 1.0$] \\
\bf{$T1$} & [$enab + 1.0, enab + 1.0$] \\
\end{tabular}
}}
\caption{Unbounded TB net model B.}
\label{fig:tw-esB}
\end{figure}

\begin{figure}
    \centering
    \subfloat[Coverability tree of B]{{\includegraphics[width=.6\textwidth]{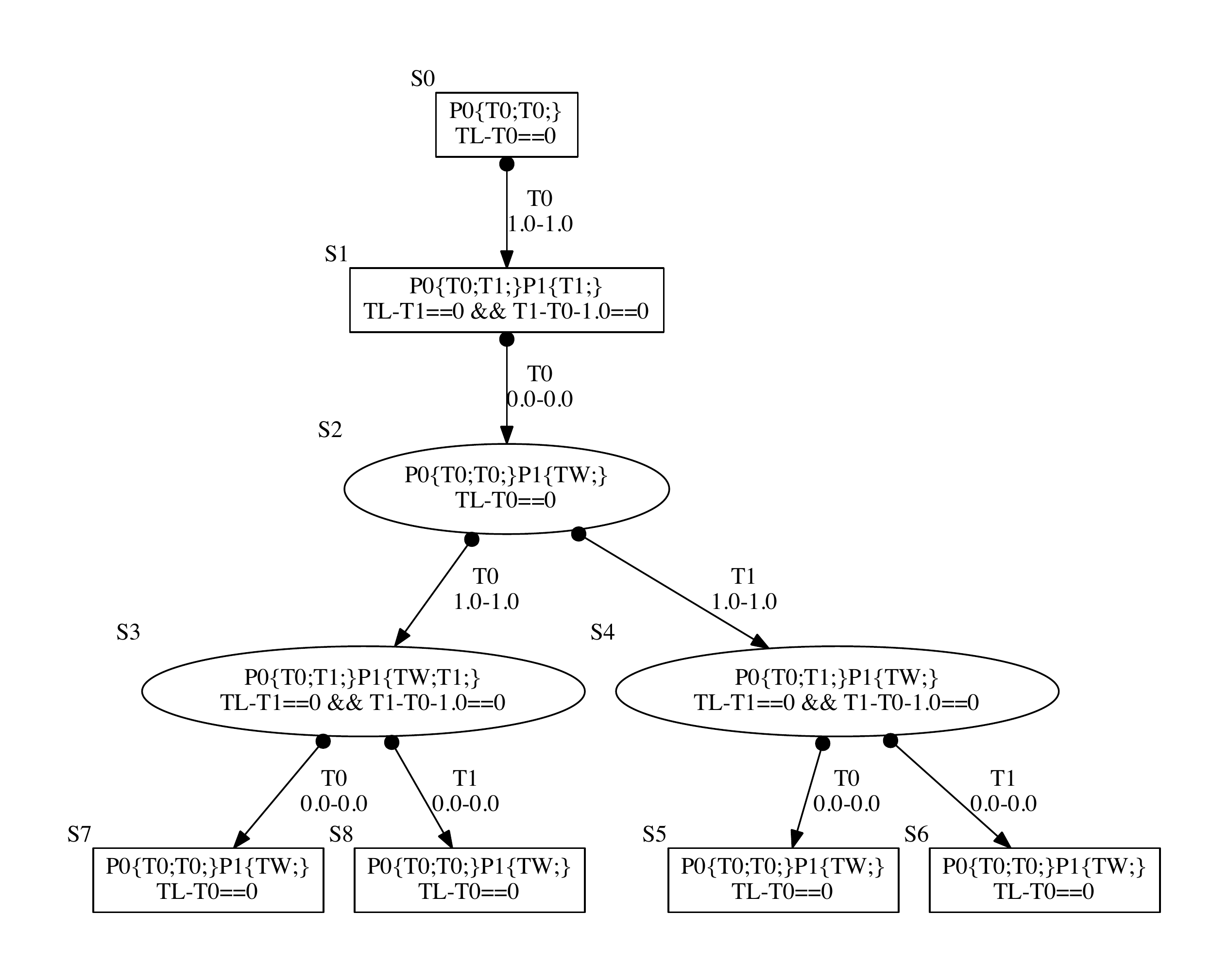} } \label{fig:B-tree} }
     \qquad
    \subfloat[Coverability graph of B]{{\includegraphics[width=.3\textwidth]{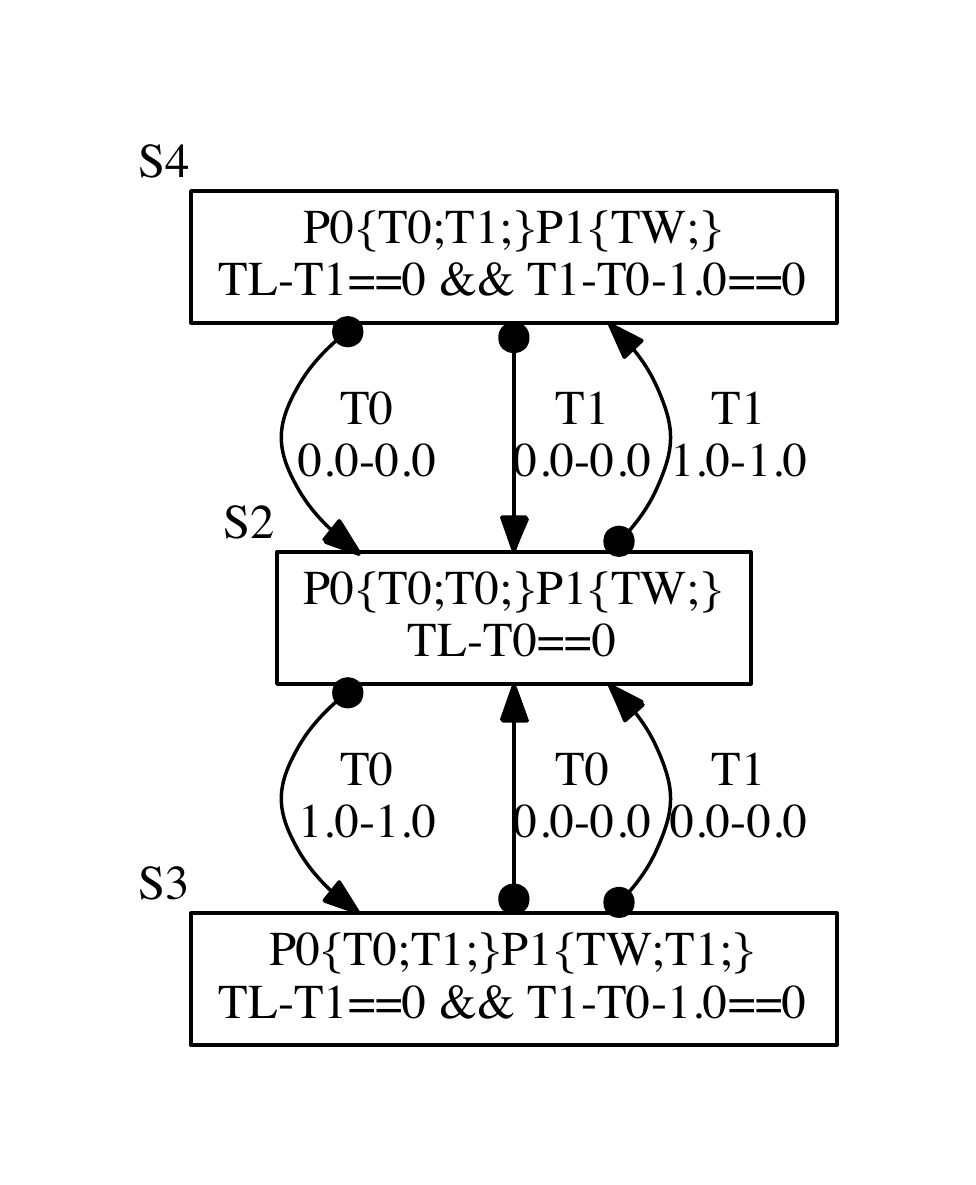} } \label{fig:B-graph} }
    \caption{Coverability tree/graph of example B.}
    \label{fig:tw-esB-tree-graph}
\end{figure}

\paragraph{Example B} This model (Figure \ref{fig:tw-esB}) is analogue to model A, except for an additional arc and a different initial marking.
It represents two synchronous tasks, where each task can either produce or consume.
An infinite buffer stores produced units. Figure \ref{fig:B-tree} and \ref{fig:B-graph} show its coverability tree and coverability graph, respectively.
It is worth noting that the firing of $T0$ from $S1$ produces an additional token into place $P1$ and because of
the recognition of both tokens of $P1$ as $TA$, the acceleration of $S2$ recognizes the $TA^\omega$ into $P1$.
Therefore, $S2$ deactivates both $S0$ and $S1$. Successors of both $S3$ and $S4$ are identified equal to $S2$.

\begin{figure}[htbf]
\centering
\includegraphics[width=.6\textwidth]{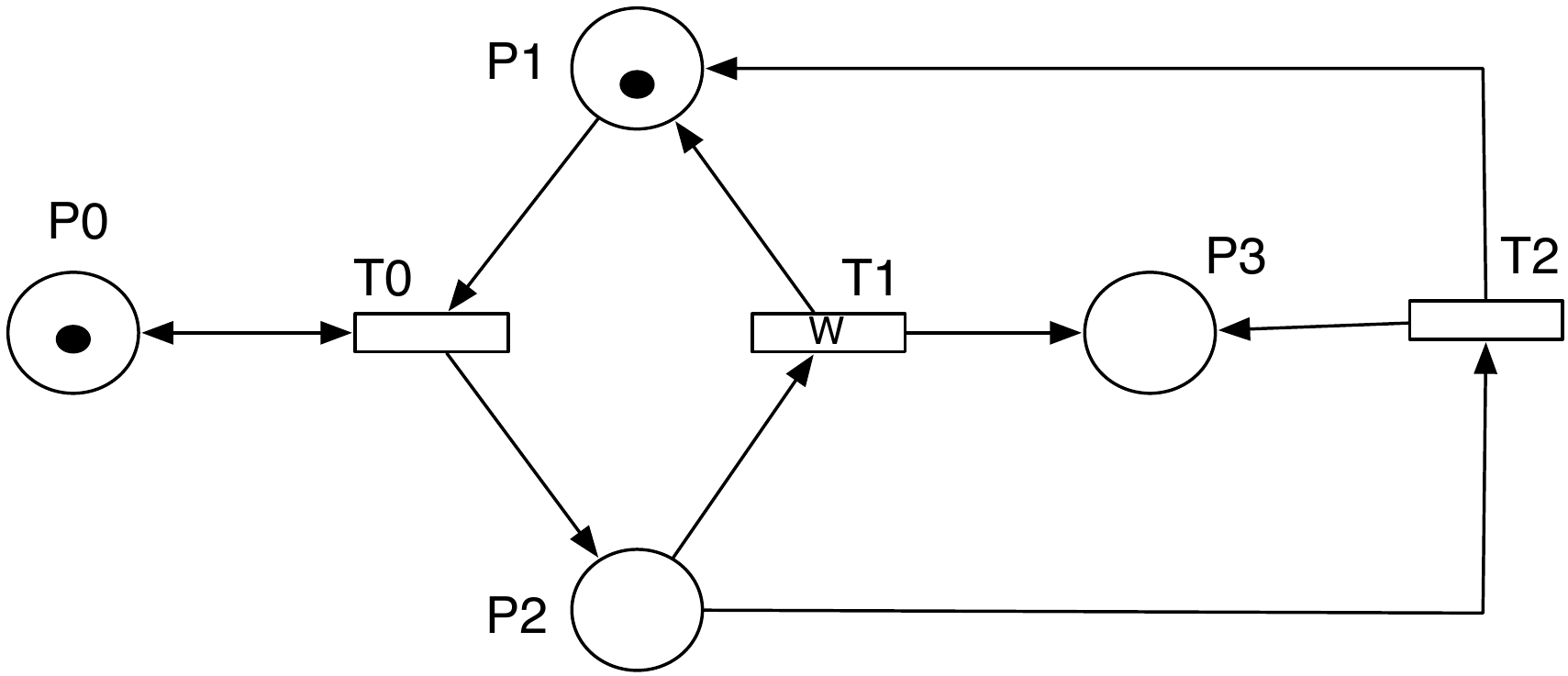}
{\small
\vskip 0.5cm
{\scriptsize
\begin{tabular}{@{} llll @{}}
\bf{Initial marking} &  $P0\{T0\} P1\{T0\}$ \\
\bf{Initial constraint} & $T0 \ge 0$ \\
 \end{tabular}
 \\
 \begin{tabular}{@{} llll @{}}
\bf{$T0$} & [$enab , P0 + 10.0$] \\
\bf{$T1$} & [$enab + 2.0, enab + 3.0$] \\
\bf{$T2$} & [$enab + 1.0, enab + 4.0$] \\
\end{tabular}
}}
\caption{Unbounded TB net model C.}
\label{fig:tw-esC}
\end{figure}

\begin{figure}
    \centering
    \subfloat[Coverability tree of C]{{\includegraphics[width=.6\textwidth]{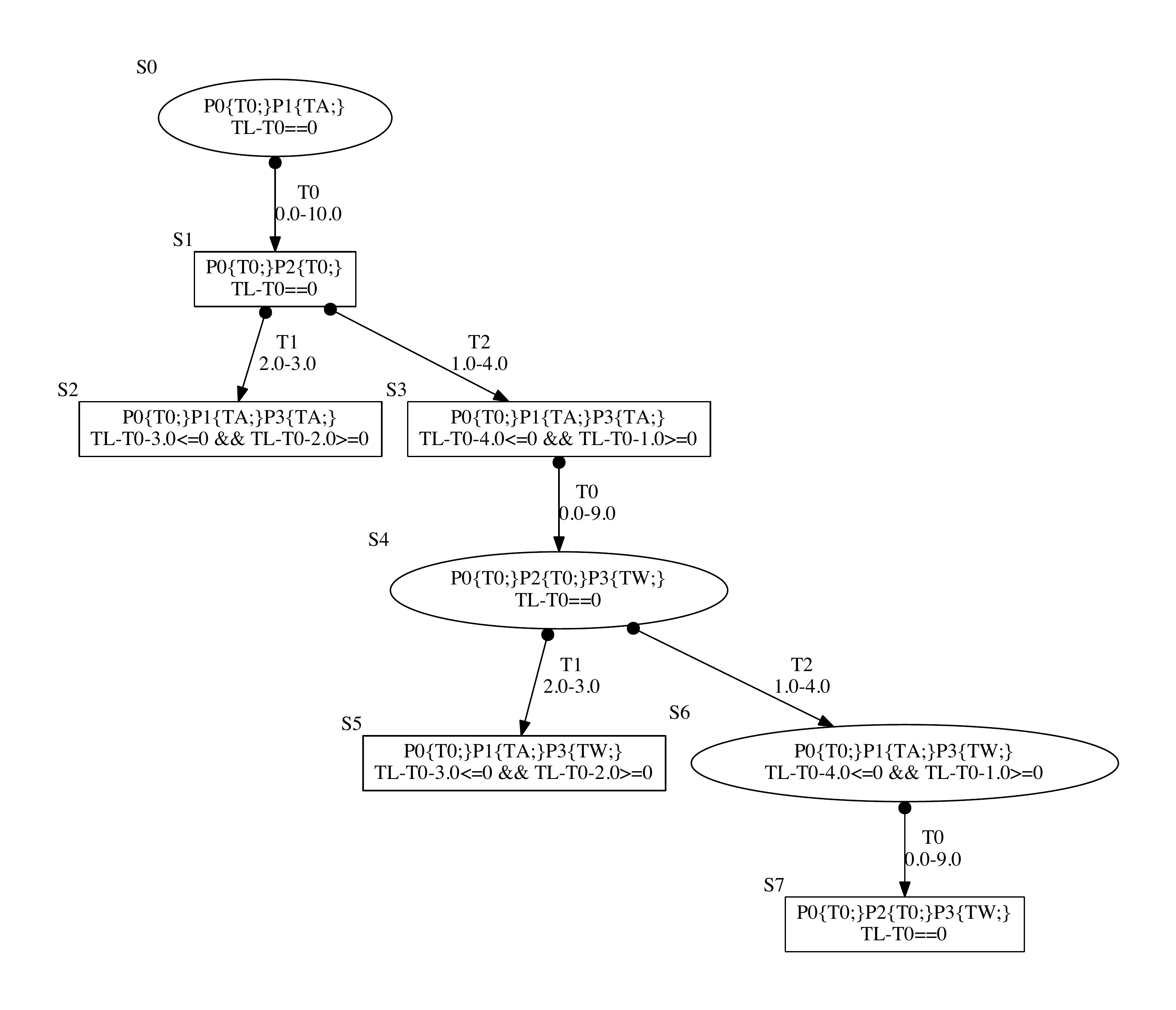} } \label{fig:C-tree} }
     \quad
    \subfloat[Coverability graph of C]{{\includegraphics[width=.3\textwidth]{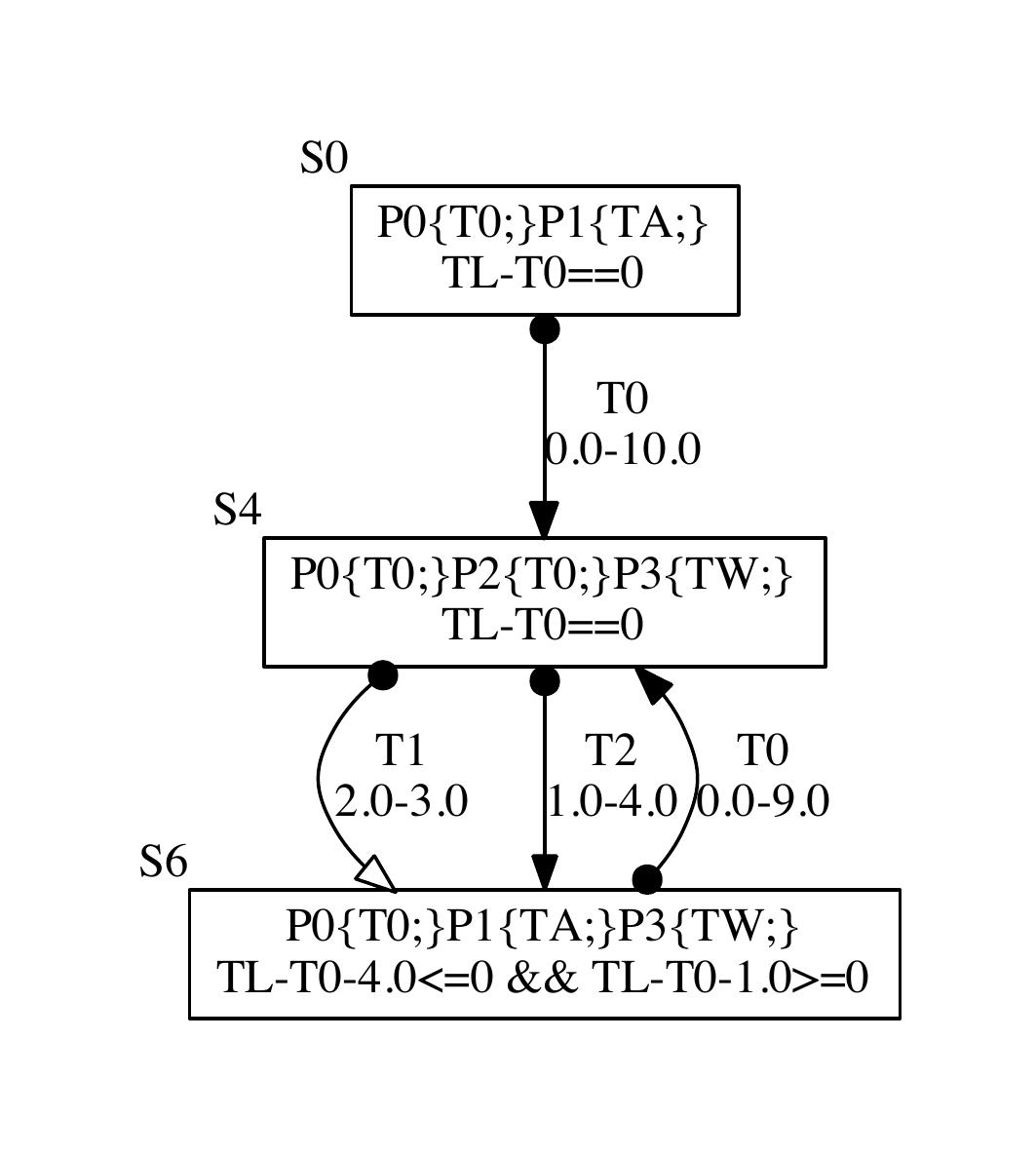} } \label{fig:C-graph} }
    \caption{Coverability tree/graph of example C.}
    \label{fig:tw-esC-tree-graph}
\end{figure}

\paragraph{Example C} This model (Figure \ref{fig:tw-esC}) represents an unbounded TB net with four places
($P0$, $P1$, $P2$, $P3$), two strong transitions ($T0$, $T2$) and a weak transition ($T3$).
Transition $T0$ acts as a sort of timer, in fact, whenever enabled, it must fire in 10 time units from its previous firing time.
Figure \ref{fig:C-tree} and \ref{fig:C-graph} show its coverability tree and coverability graph, respectively.

Concerning the current example, it is worth noting that before the introduction of $S4$, all the symbolic states were
\emph{active}. The acceleration of $S4$ leads to the recognition of a $TA^\omega$ into place $P3$, and thus
the identification of the coverage $S4 \geq S1$. This causes the deactivation of both $S1$ and its descendants $S2$ and $S3$. The successors of $S4$ are $S5$ and $S6$. In this case, since $S5 \subset S6$, $S5$ is deactivated.
Finally, $S7$ is recognized to be equal to $S4$.

\subsection{Related Work}
Concerning the reachability analysis of classic P/T nets, Karp and Miller (K\&M) introduced an algorithm for computing the \emph{minimal coverability set} (MCS) \cite{Karp69}. This algorithm builds a finite tree representation of the (potentially infinite) reachability graph of the given P/T net.
It uses acceleration techniques to collapse branches of the tree and ensure termination.
The K\&M Algorithm has been also extended to other classes of well-structured transition systems \cite{Finkel09,Finkel09II}.
Anyway, the K\&M Algorithm is not efficient in analyzing real-world examples and it often does not terminate in reasonable time. One reason is that in many cases it will compute several times a same subtree.
The MCT algorithm \cite{Finkel91} introduces clever optimizations: a new node is added to the tree only if its marking is not smaller than the marking of an existing node. Then, the tree is pruned: each node labelled with a marking that is smaller than the marking of the new node is removed together with all its successors. The idea is that a node that is not added or that is removed from the tree should be covered by the new node or one of its successors. 
However, the MCT algorithm is flawed \cite{Geeraerts07}: it computes an incomplete \emph{forward}
reachability set (i.e. all the markings reachable from the initial markings).
In \cite{Geeraerts07}, the \textsc{CoverProc} algorithm, is proposed for the computation of the MCS of a Petri net. This algorithm follows a different approach and is not based on the K\&M Algorithm.
In \cite{Reynier11}, the MP algorithm is proposed. This algorithm can be viewed as the MCT algorithm with a slightly more aggressive pruning strategy. Experimental results show that the MP algorithm is a strong improvement
over both the K\&M and the \textsc{CoverProc} algorithms.
The $TBCT$ algorithm, introduced in the current report, is somehow inspired by the MP algorithm,
and is able to construct coverability graphs of real-time systems modeled with TB nets.

For timed Petri nets (TPNs), although the set of \emph{backward} reachable states
(i.e. all the markings from which a final marking is reachable) is computable \cite{Abdulla01}, the set of \emph{forward} reachable states is in general not computable. Therefore any procedure for performing forward reachability analysis on TPNs is incomplete.
In \cite{Abdulla07}, an abstraction of the set of reachable markings of TPNs is proposed.
It introduces a symbolic representation for downward closed sets, so called region generators (i.e. the union of an infinite number of regions \cite{Alur90}).
Anyway, the termination of the forward analysis by means of this abstraction is not guaranteed.

In the current report, we addressed unbounded TB nets, which represent a much more expressive formalism for real-time systems than TPNs
(interval bounds in TB nets are linear functions of timestamps in the enabling marking,
rather than simple numerical constants). Other coverability analysis techniques for such a 
formalism, have not been proposed yet, as far as we know.

\label{sec:cg-related-work}

\subsection{Conclusion}
\label{sec:cg-remarks}
The current report introduces a coverability analysis technique able to construct a coverability tree/graph for 
unbounded TB net models. The termination of the $TBCT$ algorithm is guaranteed as long as, within the input model, tokens growing without limit, can be anonymized. This means that we are able to manage models that do not
exhibit \emph{Zeno} behavior and do not express temporal functions depending on ``infinite'' past events.
This is actually a reasonable limitation because, in general, real-world examples do not exhibit such a behavior.

\bibliographystyle{plain}
\bibliography{report}




\end{document}